\documentstyle[12pt,aaspp4,flushrt]{article} 
\begin{document}

\newcommand{\kms}{km~s$^{-1}$}
\newcommand{\subsun}{\mbox{$_{\odot}$}}

\title{AN OLD CLUSTER IN NGC 6822\altaffilmark{1}}
\vskip 6pt

\author{Judith G. Cohen\altaffilmark{2} and John P. Blakeslee\altaffilmark{2}}

\altaffiltext{1}{Based in large part on observations obtained at the
W.M. Keck Observatory, which is operated jointly by the California 
Institute of Technology and the University of California}
\altaffiltext{2}{Palomar Observatory, Mail Stop 105-24,
California Institute of Technology}

\vskip 48pt

\begin{abstract}

We present spectroscopy of two clusters in the dwarf irregular galaxy
NGC 6822.  From these we deduce an age for Cluster VII of 
$11^{+4}_{-3}$ Gyr and [Fe/H]~= $-1.95\pm0.15$ dex.  Cluster VII
appears to be an analog of the metal-poor galactic globular clusters.
Cluster VI is found to be
much younger and more metal rich, with an age of approximately 
2 Gyr.  Its derived metallicity, [Fe/H]~$\approx -1.0$ dex,
is comparable to that of the gas seen today in NGC 6822.
The existence of a metal-poor
old cluster in NGC 6822 rules out models for the
chemical evolution of this galaxy with significant
prompt initial enhancement.  We find that a star formation rate which is
constant with time and is within a factor of two of the present star formation
rate can reproduce the two points on the age-metallicity relationship
for NGC 6822 over the past 10 Gyr defined by these two clusters.
\end{abstract}


\vskip 48 pt
\centerline{\it Accepted for publication in the Astronomical Journal.}

\clearpage

\section{INTRODUCTION}

NGC 6822 is a dwarf irregular galaxy in the Local Group.  It has
several bright clusters found by Hubble (1925), with many more
cluster candidates subsequently cataloged by Hodge (1977).  While most
of them are bright blue (presumably young) clusters, one of them, Cluster VII,
is sufficiently red that it may be a true globular cluster.
The goal of this paper is to establish the age and metallicity
of this object.

The star formation history of this galaxy was studied initially
by Hodge (1980).  Marconi et al (1995), Gallart,
Aparicio \& Vilchez (1996) and Gallart et al (1996a,b) present 
more modern analyses based on CCD imaging and photometry.
Like many dwarf irregular galaxies (Hunter 1982),
star formation within NGC 6822 appears to be distributed
over isolated spatial regions at any given time, and to vary with time
at each position. Hunter (1997) reviews 
star formation processes in these small galaxies
where spiral density waves are absent.

Such studies can only examine in detail the star formation history
over the past 3~Gyr.  Older stars are fainter and are easily
swamped by the brighter younger ones.  The existence of
a genuinely old globular cluster in such a galaxy would be irrefutable
evidence for star formation more than 10 Gyr ago, an issue relevant
to the concept of bursting dwarfs introduced by 
Broadhurst, Ellis, \& Shanks (1988) and Babul \& Rees (1992),
among others, to try to explain the origin of the
excess of blue galaxies found in
deep galaxy counts.

We have recently completed an analysis of the ages and abundances of
a large sample of globular clusters in M87 (Cohen, Blakeslee \& Ryzhov 1998),
and we apply these techniques to two of the bright clusters in NGC 6822
to demonstrate that Cluster VII is indeed a genuine old globular cluster.

\section{SPECTROSCOPY OF TWO OF THE CLUSTERS IN NGC 6822}

We obtained spectra with the Low Resolution Imaging
Spectrograph (LRIS) (Oke et al 1995) on the Keck-II
telescope in September. 1997.  We used the
same instrumental configuration for NGC 6822
as in our observations of the M87 globular clusters, with a
1.5 arc-sec wide long slit replacing the multi-slit masks.
By rotating the instrument's position angle, we placed both Cluster VI and Cluster VII
of NGC 6822 in the slit.  Three exposures, each 300 sec long, were taken
followed by a flat field.

The line indices for Cluster VI and Cluster VII in NGC 6822 on the Lick system
were measured in a manner identical
to that described in our work on M87, and are listed in Table 1.
Although $H_{\beta}$ appears in absorption,
the integrated light of Cluster VI shows strong emission 
in $H_{\alpha}$, indicating
immediately that this is a young cluster.  The measurements listed
in Table 1 do not include any correction for underlying emission in
the Balmer lines.

The model grid of Worthey (1994) was used to interpret these
measurements in exactly the same way as we did for the M87 globular
clusters.  We find that Cluster VII has  $\hbox{[Fe/H]} = -1.95\pm0.15$~dex 
on the scale of Zinn (1985) and an age of $11^{+4}_{-3}$ Gyr.
This age is consistent with the mean ages of the Galactic and M87
globular clusters, which we found to be in the range of 12-15 Gyr.  

For Cluster VI, we had to extrapolate the models in order to
constrain its age and metallicity.  Doing so, we find 
$\hbox{[Fe/H]} \approx -1.0$~dex, $\hbox{age}\approx 2$~Gyr,
but these values are highly uncertain due to the large 
extrapolation, as the Worthey models only
extend down to 8~Gyr at this low metallicity and to [Fe/H] of $-$0.22
at this young age.  However, the models
clearly indicate an age less than 5 Gyr, and we expect that our 
value of 2 Gyr may be an over-estimate, as a consequence of our
ignoring any emission in the H$\beta$ line.

The metallicity we have derived for Cluster VII is consistent 
with that expected from the mean relation between globular cluster
metallicity and host galaxy luminosity (van den Bergh 1975; Harris 1991;
Perelmuter 1995).  Although the precise form of this 
relationship is still in dispute,
for dwarf galaxies with luminosities similar to that of NGC~6822,
the expected mean metallicity of any old globular cluster system is
$\langle\hbox{[Fe/H]}\rangle \approx -1.7\pm0.2$ dex, with a dispersion
of $\sim0.3$ dex for individual globulars.
Thus, it appears that Cluster VI is a compact, young open
cluster, while Cluster VII is a typical old globular cluster
(more evidence from photometry is given below).

These results support the findings of Gallart et al. (1996a), who
concluded that NGC 6822 most likely began forming stars $\sim$12~Gyr 
ago from low-metallicity gas, and has continued forming them at a slow
rate ever since.  In particular, our observations immediately eliminate
the class of ``prompt chemical enrichment'' models,
considered but deemed unlikely by these authors, wherein a
short-lived starburst of massive stars promptly enriches the gas
from primordial to a tenth of solar at very early times.
Gallart et al. state that the discovery of ``indicators of the
presence of an old, low-metallicity population, such as a horizontal
branch for example, would definitely resolve the question.''
Our observation of an old globular cluster in this galaxy
now resolves the issue.

\section{INFERRED METAL ENRICHMENT RATE}

The mean metallicity of the gas in NGC 6822 today is known from studies
of its HII regions (Lequex et al 1979; Pagel, Edmunds \& Smith 1980;
Skillman, Terlevich \& Melnick 1989). These three papers obtain
consistent results of O/H $\sim$ 0.15 the solar value.
In old metal poor stars in our own galaxy, O/H is enhanced over
Fe/H by a factor of $\sim$2 (Wheeler, Sneden \& Truran 1989).  
If we assume this is also the case in NGC 6822, then the current [Fe/H]
of its gas is $\approx -1.1$ dex, consistent with the very
young age and similar metallicity for Cluster VI,
while Cluster VII has a metallicity some $\sim$7 times lower. 
This difference represents the accumulated
metal formation by generations of stars in NGC 6822.

Hodge (1993) reports a current star formation rate (SFR) of
0.021~$M_\odot$~yr$^{-1}$ for NGC 6822 based on its global H$\alpha$
luminosity. From an estimate of the total stellar mass of this galaxy,
he concludes that the time-averaged SFR is very close to the current
one; in particular, if NGC 6822 has been forming stars for 12~Gyr,
the inferred mean SFR is $\sim\,$0.028~$M_\odot$~yr$^{-1}$.
Gallart et al.\ (1996b) find a SFR of 0.04~$M_\odot$~yr$^{-1}$
over the past $\sim\,$200~Myr, which they interpret as a stochastic
enhancement of roughly a factor of 2 over the mean SFR.
They point to Hodge's current value as perhaps indicating that the SFR
has once again decreased.  In any case, the data indicate that NGC 6822
has been steadily forming stars at a very modest rate of
0.02--0.04~$M_\odot$~yr$^{-1}$ throughout its lifetime.

We can ask if the metal enrichment rate inferred from our data
is consistent with a constant star formation rate approximately 
equal to the observed one.  A naive ``closed-box'' model of chemical
enrichment (see, for example, Binney \& Tremaine 1987)
predicts that the metallicity of the interstellar gas will increase
with time as:
$$Z(t) = Z_0 - p \times \hbox{ln}\left[ 1 - {M_s(t) \over M_{s+g}}\right],$$
where $Z_0$ is the initial metal fraction, $M_s(t)$ is the mass in
stars at time $t$, $M_{s+g}$ is the constant total mass in stars and
gas, and $p$ is the ``yield parameter'', the fraction of input mass 
that gets returned to the interstellar gas in the form of heavy elements.  
The mean SFR is  $\langle \dot{M}_s\rangle \equiv M_s(t)/\Delta t$.
The value of $p$ is fairly uncertain, but a reasonable
estimate is $p = 0.005\pm0.003$ (e.g., Rana 1991;
Edmunds \& Pagel 1984). Adopting values of 
$\Delta Z \approx 0.0015 \pm 0.0007$, $\Delta t \approx 10\pm3$~Gyr
for the difference in metallicity and age between Clusters VI and VII, 
and taking $M_{s+g}$ from Gallart et al., 
we find a mean SFR in NGC 6822 of 
$\langle \dot{M}_s\rangle = 0.04^{+0.04}_{-0.03}$~$M_\odot$~yr$^{-1}$,
consistent with other estimates and with the currently observed SFR.
Of course, if infall of primordial gas was important, than
in order to produce the same amount of metal enrichment, the
true time-averaged SFR would have to be higher than our naive estimate.

\section{PHOTOMETRY AND RADIAL VELOCITY OF CLUSTER VII}

The distance modulus for NGC 6822 comes from
its Cepheids, originally studied by Kayser (1967), with
a recent reanalysis by Visvanathan (1989).  We adopt the reddening
and distance modulus for this galaxy of Gallart, Aparicio \& Vilchez (1996).
The observed colors for Cluster VII are $(U-B)$ = 0.30 mag and $(B-V)$ = 0.85 mag, 
with $V$ = 15.78 mag, while those
of Cluster VI are $\sim$0.5 mag bluer in both $(B-V)$ and $(U-B)$
(van den Bergh \& Humphreys 1979, Wilson 1992), as expected.  
The luminosity of Cluster VII is then $M_V = -8.45$ mag 
($L = 1.9 \times 10^5 L\subsun$.
This corresponds to a luminosity somewhat brighter than that
of the peak of the galactic globular cluster luminosity function at 
$M_V = -7.35$ mag (Zinn 1985).  

The unreddened colors for Cluster VII are $(B-V)_0$ = 0.61 mag,
$(U-B)_0 = 0.11$ mag.  These agree to within the uncertainties of
measurement with the colors of metal poor galactic globular clusters
such as M92, M13 and M3 as tabulated by Reed, Hesser \& Shawl (1988).
Wilson (1992) arrived at the opposite conclusion, and asserted that
Cluster VII is similar to the intermediate age clusters of the Magellanic Clouds, but she was using
a reddening value substantially larger than the one adopted here.

The distribution of the HI gas in the central region of NGC 6822 has 
been mapped by Gottesman \& Weliachow (1977), who see a rotating disk, a
pattern typical of such galaxies.  The observed heliocentric
radial velocity from our LRIS spectra of Cluster VII is $-52$ \kms, which corresponds
closely to the $v_r$ expected from the HI, so there is
no evidence for a ``halo'' on kinematic grounds.

\section{SUMMARY}

We have established from analysis of its spectral features that
Cluster VII in NGC 6822 appears to be a metal-poor old globular cluster
with an age of $11^{+4}_{-3}$ Gyr and [Fe/H] = $-1.95\pm0.15$ dex. 
Cluster VI in NGC 6822 is found to be much younger and more metal-rich
with  an age of approximately 2 Gyr.
The metallicity derived for Cluster VI, [Fe/H] $\approx -1.0$ dex,
is comparable to that of the gas
in NGC 6822 at the present epoch. The colors of
these two objects are consistent with this interpretation.
This eliminates models of the chemical evolution of NGC 6822 that invoke
strong prompt initial enrichment
and has implications for the idea that bursting
dwarf galaxies are the origin of the excess numbers of faint blue objects found
in deep galaxy counts.  For instance, these results do not
support the suggestion
(Broadhurst et al.\ 1988)
that occasional strong bursts of star formation in fairly
isolated, gas-rich dwarfs are responsible for the excess faint counts,
although late-forming dwarf ellipticals and spheroidals which deplete their
gas supply in a strong initial burst (Babul \& Rees 1992) are still a possibility.
A star formation rate that is constant
in time and within a factor of two of the present value can reproduce
the two known points provided by 
Clusters VI and VII on the age-metallicity relationship for NGC 6822.

\acknowledgements
The entire Keck/LRIS user community owes a huge
debt to Jerry Nelson, Gerry Smith, Bev Oke, and many other people who
have worked to make the Keck Telescope and LRIS a reality.  We are grateful to
the W. M. Keck Foundation, and particularly its late president, Howard
Keck, for the vision to fund the construction of the W. M. Keck
Observatory.  

JGC is grateful to NSF grant AST96-16729 for support.
JPB is grateful to the Sherman Fairchild Foundation for support.

\begin{deluxetable}{lrrrrrrrrr}
\tablewidth{0pt}
\footnotesize
\tablecaption{Indices on the Lick System for Two Clusters in NGC 6822}
\tablehead{
\colhead{ID} & \colhead{H$\beta$} & \colhead{Mg$_1$} & \colhead{Mg$_2$} & 
\colhead{Mg $b$} & \colhead{Fe5270} & \colhead{Fe5335} & \colhead{NaD} &
\colhead{TiO$_1$} & \colhead{H$\alpha$} \nl
\colhead{} & \colhead{(\AA)} & \colhead{(mag)} & \colhead{(mag)} & 
\colhead{(\AA)} & \colhead{(\AA)} & \colhead{(\AA)} & \colhead{(\AA)} &
\colhead{(mag)} & \colhead{(\AA)}
}
\startdata
Cluster VI & 3.75 & 0.017 & $-$0.002 & 0.308 & 0.682 & 0.610 &
      1.02 & 0.004 & $-$6.45 \nl
Cluster VII & 2.70 & 0.032 & 0.004 & 0.581 & 0.731 & 0.810 & 
      1.12 & 0.005 & 2.27 \nl
\enddata
\vspace{1.5 truecm}
\end{deluxetable}

%
%

\end{document}